\documentclass[12pt]{iopart}

\usepackage{color}
\usepackage[normalem]{ulem}
\usepackage{bm}
\usepackage{graphicx}

\begin{document}

\title[STM study of exfoliated bP]{STM Study of Exfoliated Few Layer Black Phosphorus Annealed in Ultrahigh Vacuum}

\author{Abhishek Kumar$^1$, F. Telesio$^1$, S. Forti$^2$, A. Al-Temimy$^2$, C. Coletti$^2$, M. Serrano-Ruiz$^3$, M. Caporali$^3$, M. Peruzzini$^3$, F. Beltram$^1$, and S. Heun$^1$}

\address{$^1$ NEST, Istituto Nanoscienze-CNR and Scuola Normale Superiore, Piazza San Silvestro 12, 56127 Pisa, Italy}
\address{$^2$ Center for Nanotechnology Innovation @ NEST, Istituto Italiano di Tecnologia, Piazza San Silvestro 12, 56127 Pisa, Italy}
\address{$^3$ CNR-ICCOM, Via Madonna del Piano 10, 50019 Sesto Fiorentino, Italy}
\ead{stefan.heun@nano.cnr.it}

\vspace{10pt}
\begin{indented}
\item[]July 2018
\end{indented}

\begin{abstract}
Black Phosphorus (bP) has emerged as an interesting addition to the category of two-dimensional materials. Surface-science studies on this material are of great interest, but they are hampered by bP's high reactivity to oxygen and water, a major challenge to scanning tunneling microscopy (STM) experiments. As a consequence, the large majority of these studies were performed by cleaving a bulk crystal \textit{in situ}. Here we present a study of surface modifications on exfoliated bP flakes upon consecutive annealing steps, up to 550~$^\circ$C, well above the sublimation temperature of bP. In particular, our attention is focused on the temperature range 375~$^\circ$C -- 400~$^\circ$C, when sublimation starts, and a controlled desorption from the surface occurs alongside with the formation of characteristic well-aligned craters. There is an open debate in the literature about the crystallographic orientation of these craters, whether they align along the zigzag or the armchair direction. Thanks to the atomic resolution provided by STM, we are able to identify the orientation of the craters with respect to the bP crystal: the long axis of the craters is aligned along the zigzag direction of bP. This allows us to solve the controversy, and, moreover, to provide insight in the underlying desorption mechanism leading to crater formation.
\end{abstract}

\section{Introduction}

The field of two dimensional (2D) materials was first opened by Geim and Novoselov in 2004, when they exfoliated graphite \cite{Novoselov2004} and studied graphene \cite{Novoselov2005, Novoselov2007}, the one--atom--thick layer of graphite, a work for which they were awarded the Nobel prize in 2010. This led to the discovery of many other layered materials with a wide range of properties, from conducting semi-metal graphene to semiconducting direct-band gap transition-metal dichalcogenides (TMDCs), and to insulating hexagonal Boron Nitride (h-BN) \cite{Geim2013}. 2D-Xenes \cite{Molle2017}, monolayered sheets of group IV-A elements like Silicon, Germanium, and Tin, called Silicene \cite{Vogt2012,Feng2012,Meng2013}, Germanene \cite{Davila2014}, and Stanene \cite{Zhu2015}, respectively, have been realized experimentally, and found to have interesting properties like large, tunable band gap and robust quantum spin Hall states, suitable for applications in the field of room temperature 2D topological insulators \cite{Xu2013}, anomalous Seebeck effect devices \cite{Xu2014}, and other spintronic applications \cite{Rachel2014}. Borophene, a monolayer sheet of Boron, has been synthesized and reported to have a highly anisotropic metallic character \cite{Mannix2015, Feng2016}.

Phosphorene is a recent addition to this group. It has been exfoliated and realized as a 2D material only in 2014 \cite{Liu2014}. Phosphorene is a single sheet of phosphorus atoms, obtained by exfoliation from the orthorombic black phosphorus crystal. By convention, the monoatomic layer is assigned to the a-c crystallographic plane, with $\boldsymbol{a} = [100]$ as zigzag and $\boldsymbol{c} = [001]$ as armchair directions. Each P atom is sp$^3$ hybridized \cite{Li2015} and forms covalent bonds with three neighboring P atoms, leaving one lone pair. This leads to a puckering of the monoatomic layers due to the repulsion of the P--P bonds by these lone electron pairs \cite{Asahina1982,Morita1981}. The direct band gap of bP can be tuned from $\sim 0.3 - 2.0$~eV by the number of layers (from bulk to monolayer) \cite{Das2014}, and this electronic property has allowed bP to occupy a very important spot between the zero-band gap graphene and the large-band gap TMDCs \cite{Qiao2014,Liu2014}. Moreover, the presence of a crystallographic anisotropy imparts anisotropy on physical properties like mobility \cite{Liu2016}, effective mass \cite{Hu2015}, magneto--transport \cite{Hemsworth2016}, and plasmon resonance \cite{Serrano2016}, thus offering new opportunities for its use in electronic, optical, and thermoelectric devices.

However, bP suffers from a thickness--dependent surface instability in air, because of a combined effect of oxygen, water, and light \cite{Favron2015, Huang2016, Luo2016}, which represents a major drawback for practical device applications. It is therefore very important to investigate and thoroughly understand the bP surface properties. Earliest work in this regard was performed in 1992, probing the bP surface using scanning tunneling microscopy (STM) in air and showing the first atomic resolution images of this material \cite{Yau1992}. The zigzag pattern corresponding to the upper atomic plane of the bP top layer was clearly identified in this work. However, the need to routinely cleave the material before imaging, and still observing surface degradation and modification while scanning, imposed the need of inert atmosphere for such experiments. Later STM work \cite{Zhang2009,Liang2014,Kiraly2017,Riffle2017, Qiu2017,Gui2017} under ultra--high vacuum (UHV) conditions provided clearer and more stable atomic resolution images. The measured band gaps correspond to the predicted values, and phenomena like edge reconstruction, single atomic vacancies, impurities, and hydrogenation and phosphorization of oxygenated bP were explored \cite{Zhang2009,Liang2014,Kiraly2017,Riffle2017, Qiu2017,Gui2017}.

All these studies were performed on bulk bP crystals cleaved inside a UHV chamber, while from a 2D application point of view, exploring exfoliated thin flakes is more relevant. In this regard, no STM work investigating the surface behavior of exfoliated bP flakes was reported, and only one recent paper, focused on spectroscopic properties, showed an atomic-resolution image of an exfoliated bP flake \cite{Liu2017a}. The thermal stability and the temperature-dependent surface degradation of thin flakes of bP also needs to be well understood, since most synthesis methods and conventional device processing require high-temperature treatments. 

Here, we present an STM investigation of surface mor\-pho\-lo\-gical changes of $\sim 30$~nm--thin exfoliated bP flakes under UHV conditions as driven by temperature variation. We found that an annealing temperature of 300~$^\circ$C to 350~$^\circ$C is high enough to clean the surface from oxides and water, while still not altering the original surface of the bP flakes. Upon annealing to higher temperatures (375~$^\circ$C - 400~$^\circ$C), we observe crater formation on the surface, consistent with what was reported in earlier studies using transmission electron microscopy (TEM) \cite{Liu2015} and low energy electron microscopy (LEEM) \cite{Fortin-Deschenes2016}. Upon further annealing at 550~$^\circ$C - 600~$^\circ$C, we observe surface roughening due to sublimation of phosphorus atoms, in agreement with other studies \cite{Liu2015}. 

The mechanism of bP desorption with temperature and on different crystallographic surfaces of bP \cite{YOO2018} is an important aspect of technological relevance that needs to be understood well, as it will allow to control and tune bP flake thickness via annealing \cite{Robbins2017,Luo2017, Lin2017} and to face the complex challenge of chemical vapor deposition (CVD) growth of bP \cite{Lin2017}. Furthermore, since any chemical functionalization of bP will mainly occur at the step edges of the surface, it is of utmost importance to know the detailed geometry and orientation of the steps that form on the surface.

Desorption of bP was investigated by Liu et al.~\cite{Liu2015} and Fortin-Deschenes et al.~\cite{Fortin-Deschenes2016}. In support to their experimental evidence, different theoretical models were proposed and discussed by them, corresponding to atomic P and molecular P$_2$ desorption, respectively. Both models lead to an\-isotrop\-ic eye-shaped craters and an alignment of the long axis of these craters to a crystallographic direction. They differ, however, in their conclusions: (i) atomic P desorption leads to craters aligned along [001] (armchair) direction \cite{Liu2015}; (ii) molecular P$_2$ desorption leads to craters aligned along [100] (zigzag) direction \cite{Fortin-Deschenes2016}. These studies were based on measured diffraction patterns, using TEM and LEEM, respectively, but both studies lack atomic--resolution imaging. Thanks to our atomic resolution, here we are able to resolve this debate and show that the orientation of the craters is aligned to the lattice crystallographic [100] or zigzag direction.

\section{Results}

In our experiments, we used epitaxial monolayer graphene grown on the Si--face of 6H N-doped semiconducting Silicon Carbide [SiC(0001)] as a substrate. bP was exfoliated on this substrate in an inert atmosphere in order to preserve the samples from degradation during exfoliation. Being conductive, graphene provides the required channel for the tunneling current to flow, which makes these measurements possible without the need for additional contacts to the bP flakes, thus avoiding fabrication steps that might contaminate the sample. Overall, the surface of the graphene samples appears flat, with SiC terraces 700 to 800~nm wide, which allows clear identification of bP flakes. Step heights are multiples of single SiC bilayers, because of step bunching related to graphene growth \cite{Sprinkle2010}.

\begin{figure}[tbp]
   \centering
   \includegraphics[width=0.65\textwidth]{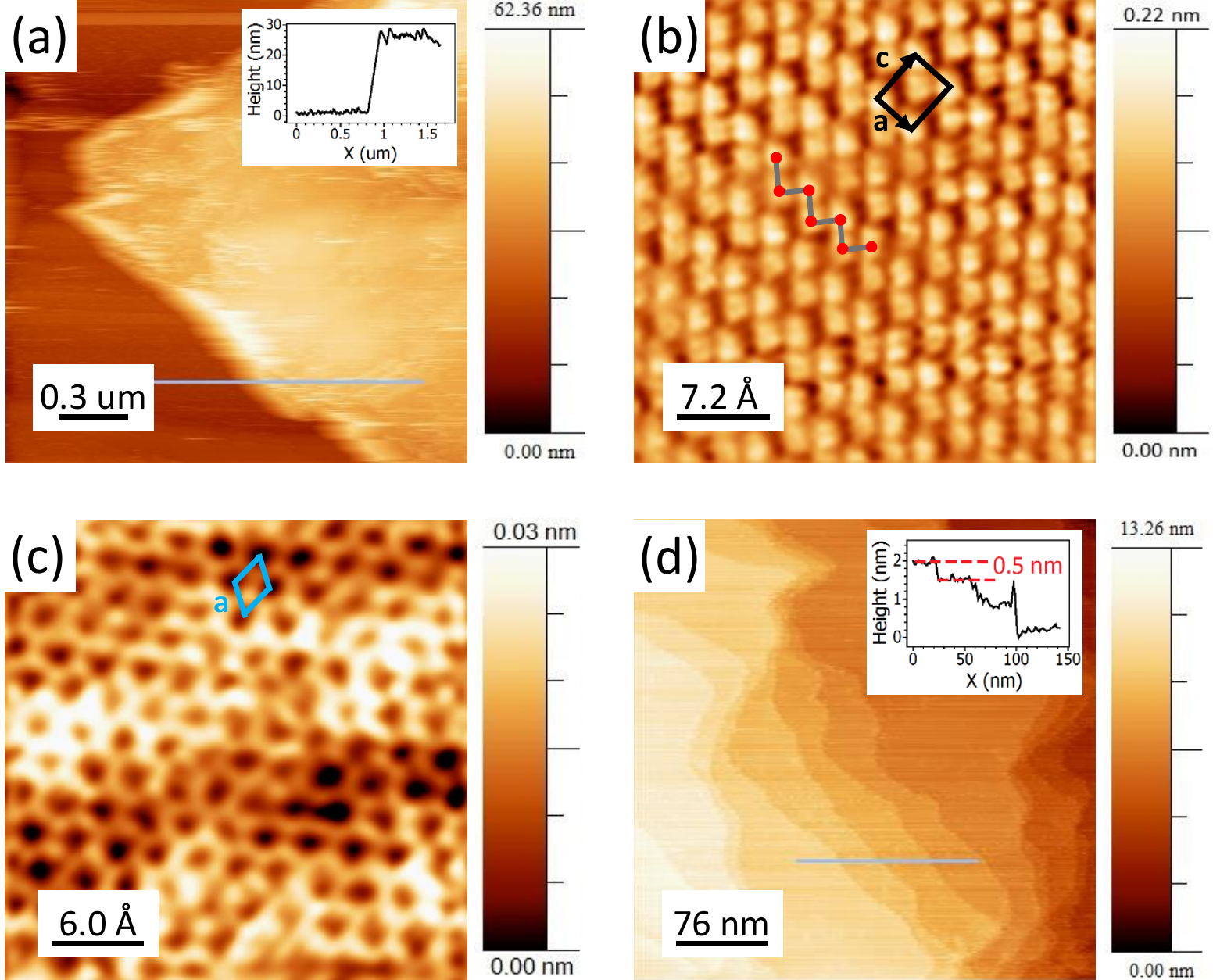}
   \caption{\label{Fig1}STM images identifying bP flakes and graphene substrate. (a) STM image showing on the right a bP flake $\sim 25$~nm high above the graphene substrate on the left. The inset shows the height profile across the line shown in the STM image. Scan size: 2~$\mu$m $\times$ 2~$\mu$m, imaging parameters: (0.7 V, 300 pA). Annealing conditions: 200~$^\circ$C, 2~h. (b) Atomic resolution image obtained on bP at room temperature showing the zigzag pattern with unit cell parameters $a = (3.45 \pm 0.43)$ {\AA} and $c = (4.40 \pm 0.12)$ {\AA}. Scan size: 3.6~nm $\times$ 3.6~nm, imaging parameters: (0.7 V, 25~pA). Anealing conditions: 400~$^\circ$C, 10~min. (c) Atomic resolution image on graphene. Unit cell indicated. Scan size: 3~nm $\times$ 3~nm, imaging parameters: (0.1 V, 157~pA). Annealing conditions: 450~$^\circ$C, 2~h. (d) Steps measured on a bP flake. The height profile in the inset, taken along the line in the STM image, shows a step height of 0.5~nm, consistent with monolayer bP steps. Scan size: 380~nm $\times$ 380~nm, imaging parameters: (2 V, 100~pA). Annealing conditions: 400~$^\circ$C, 10~min.}
\end{figure}

In STM imaging, the bP flakes are identified as plateau-like structures --- several nanometers high, and flat. A typical flake is $\sim 1$~$\mu$m$^2$ in area and $\sim 30$~nm high (see Fig.~S1 in the Supplementary Data), which gives a first clue towards identification of bP flakes. Figure 1(a) shows a typical sample, with graphene on the left and a $\sim 25$~nm--high flake on the right. Flake height is extracted from the cross--section shown in the inset. Zooming in and measuring atomic-resolution images confirms the nature of these structures. Figure~1(b) shows atomic resolution on bP, with the zigzag pattern well visible and unit cell parameters $a = (3.45 \pm 0.43)$~{\AA} and $c = (4.40 \pm 0.12)$~{\AA}, close to literature values \cite{Brown1965,Cartz1979,Morita1986,Zhang2009,Brent2014}. Figure 1(d) shows steps on a bP flake that are 0.5~nm high or multiples of this value, consistent with literature values for the interlayer distance in bP \cite{Morita1981}. Figure 1(c) shows atomic resolution on graphene, with the typical hexagonal pattern and a unit cell parameter $a = (2.46 \pm 0.22)$ \AA, consistent with the reported value \cite{Hass2008,Mishra2016}. Thus, we can clearly confirm the chemical identity of the materials under investigation, both for bP and for graphene. This demonstrates that we are able to prepare samples with exfoliated bP flakes suitable for STM investigations, coping with the problem of the high surface reactivity commonly degrading the flakes. We are also able to reliably distinguish the bP flakes from the substrate and can zoom in to detect the surface morphology of these flakes at high resolution, which allows us to study surface behavior with temperature.

\subsection{Annealing experiments}

Once the bP flakes were identified and characterized, we started annealing experiments to observe the surface morphological changes with temperature. We annealed the samples in steps of 50~$^\circ$C, starting from 150~$^\circ$C, and performed STM imaging after each step. A complete set of representative images is given in the Supplementary Data, Fig.~S2.

\begin{figure*}[tbp]
   \centering
   \includegraphics[width=0.65\textwidth]{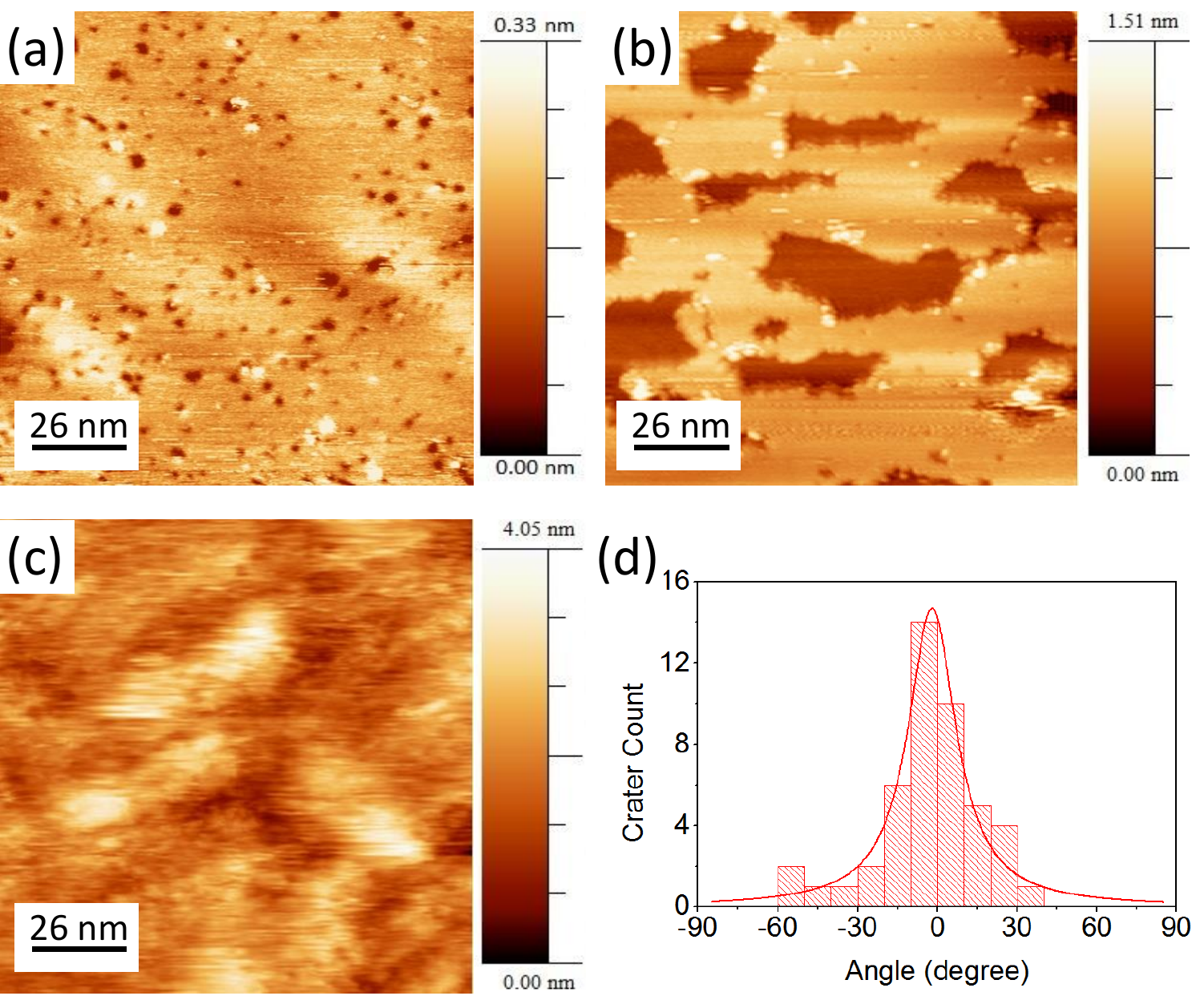}
   \caption{\label{Fig2}Annealing experiment. STM images of 130~nm $\times$ 130~nm scan area with annealing temperature, annealing time, and scanning parameters: (a) 300~$^\circ$C, 2~h, (1.2 V, 100~pA), (b) 400~$^\circ$C, 10~min, (2 V, 100~pA), (c) 500~$^\circ$C, 2~h, (1.2 V, 100~pA), respectively. (d) Histogram showing the distribution of crater angle orientation in one of the areas measured after annealing at 375~$^\circ$C, 10~min. 0$^\circ$ corresponds to the horizontal axis in the image.}
\end{figure*}

Below 300~$^\circ$C, we did not achieve atomic resolution, since the temperature was likely too low to suitably clean the samples. However, after annealing the samples at 300~$^\circ$C - 350~$^\circ$C, we obtained clear atomic resolution at room temperature, thus showing that this treatment yields an oxide--free and clean surface for STM investigations. We also observed the presence of some defects on the surface, appearing as dark and bright spots, as for example shown in the 130~nm $\times$ 130~nm scan-area image shown in Fig.~\ref{Fig2}(a). Similar defects were reported in few papers at the moment. Those defects have been ascribed to vacancies \cite{Kiraly2017, Riffle2017}, oxygen trapped in the lattice \cite{Gui2017}, or to Sn impurities related to the bP crystal growth procedure \cite{Qiu2017}. Since a systematic investigation of the nature of these defects is not the purpose of our present study, we do not provide a clear identification of these defects. We however underline that the size of these defects, 5~nm to 8~nm, is in agreement with the reported literature values \cite{Kiraly2017, Gui2017, Riffle2017, Qiu2017}.

Upon annealing to higher temperatures, at 375~$^\circ$C - 400~$^\circ$C, phosphorus atoms started desorbing, which left craters behind at the surface, as shown in Fig.~\ref{Fig2}(b). It is evident just by their appearance that these craters have a preferred direction of alignment, which is not random. In a statistical analysis, we measured the angle of these craters with respect to the horizontal image direction and plotted them in histograms. We analyzed 3 flakes following different annealing conditions: [(375~$^\circ$C, 10~min), (400~$^\circ$C, 5~min), and (400~$^\circ$C, 10~min)], comprising about 80 craters (see Fig.~S3 in Supplementary Data). Figure~\ref{Fig2}(d) shows the histogram plot with the largest number of craters in the image, fitted with a Lorentzian distribution function with a peak (mean angle) at $(-2.0 \pm 0.6)^\circ$ and a FWHM (angle scatter) of $(22.2 \pm 1.9)^\circ$. Thus, most of the craters point in one preferred direction, with none pointing along the orthogonal direction, as underlined by the absence of even a single crater at 90$^\circ$ in the histogram plot. To confirm that the craters were not formed by the tip while scanning, we scanned one of the locations with craters several times at 0$^\circ$ and 90$^\circ$ to the horizontal, and observed the craters being unaltered by changing the scan direction (see Fig.~S4 in Supplementary Data). As already mentioned, similar craters were reported before \cite{Fortin-Deschenes2016, Liu2015}, but here we can resolve craters that are one order of magnitude smaller in size than those previously observed. The growth of larger craters could be the result of the coalescence of several smaller craters like those shown by our experiments. Thanks to the high resolution provided by STM, we can therefore resolve the initial stages of crater formation at 375~$^\circ$C - 400~$^\circ$C, and show that even these have a preferred orientation.

When annealing at higher temperatures (450~$^\circ$C - 500~$^\circ$C), as we go much above sublimation temperature, the surface became rough, as shown in Fig.~\ref{Fig2}(c). The RMS roughness increased from about 0.3~\AA{} to 6~\AA. At 550~$^\circ$C - 600~$^\circ$C, most of the flakes have desorbed. The flake density, which in the beginning of the experiment was approximately 0.01 flakes per $\mu$m$^2$ (one flake in 100~$\mu$m$^2$), decreased by more than one order of magnitude.

\subsection{Crater Orientation}

\begin{figure*}[tbp]
   \centering
   \includegraphics[width=\textwidth]{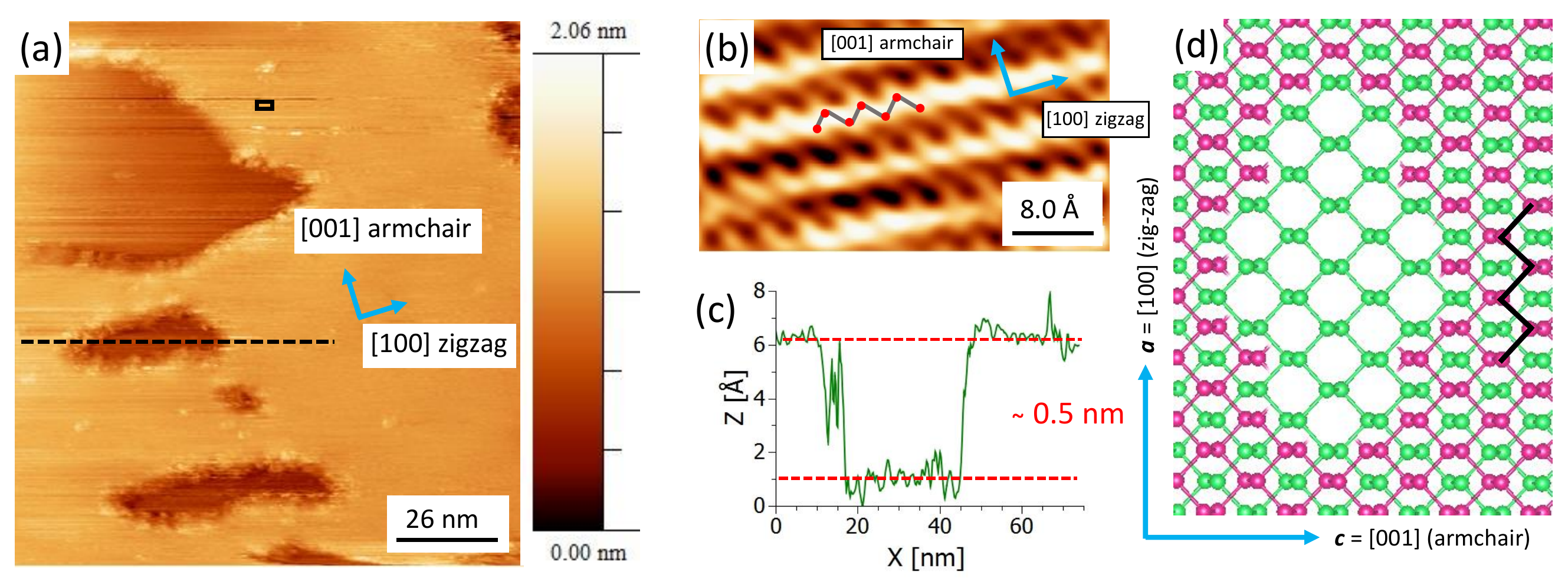}
   \caption{\label{Fig3}Crystallographic direction of crater alignment. (a) STM image of a 130~nm $\times$ 130~nm scan area, showing aligned craters on bP after annealing at 400~$^\circ$C for 2~h; scanning parameters: (1.2 V, 100~pA). (b) Atomic resolution image obtained after zooming into the region marked in (a), providing information of the crystallographic directions of the bP flake; scanning parameters: (1.2 V, 100~pA). (c) Height profile across the crater along the dashed line in (a), showing $\sim 0.5$~nm step height, compatible with monolayer desorption. (d) Schematics showing crater formation due to bP desorption - bottom layer (green) visible under desorbing top layer (red). Crystallographic direction is denoted by black zigzag drawn along [100] direction.}
\end{figure*}

Figure~\ref{Fig3} shows a bP surface similar to the one shown in Fig.~\ref{Fig2}(b), obtained by annealing the sample at 400~$^\circ$C for 2~h. Figure~\ref{Fig3}(a) shows a 130~nm $\times$  130~nm image of this surface. We can see asymmetric craters, with their long axes aligned. The height profile along the line in Fig.~\ref{Fig3}(a) across the crater is shown in Fig.~\ref{Fig3}(c). The depth of the crater of $\sim 0.5$~nm indicates monolayer desorption. In order to determine the crystallographic alignment of these craters, we zoomed in for atomic resolution. Figure~\ref{Fig3}(b) shows a zoomed--in image acquired in the region indicated by the black rectangle in Fig.~\ref{Fig3}(a). The zigzag and armchair directions are clearly identified in Fig.~\ref{Fig3}(b). Thus, we can attribute the long axis of the craters in Fig.~\ref{Fig3}(a) to the [100] (zigzag) direction. Together with our result regarding the preferential orientation of the craters, we can conclude that the asymmetric craters form owing to phosphorus desorption during annealing, and they tend to align along the zigzag [100] crystallographic direction, in agreement with the study by Fortin-Deschenes et al.~\cite{Fortin-Deschenes2016}.

\section{Discussion}

With respect to the underlying mechanism of phosphorus desorption behind the crater formation, there are two competing models: (i) atomic P desorption leading to craters aligned along [001] (armchair) direction \cite{Liu2015}; (ii) molecular P$_2$ desorption leading to craters aligned along [100] (zigzag) direction \cite{Fortin-Deschenes2016}. For the case of monoatomic P desorption, Liu et al.~\cite{Liu2015} proposed a simplistic model of crater formation by propagation of cracks from a monoatomic vacancy. Starting from a single missing atom on the bP surface, the unsaturated atoms around the vacancy (i.e.~P atoms forming less than three bonds) are removed step by step. The assumed order of removal of different types of unsaturated P atoms is \cite{Liu2015}: (a) P atoms with only one bond desorb first. This is followed by (b) P atoms with two in-plane bonds which desorb next, and finally (c) desorption of P atoms with one in-plane bond and one out-of-plane bond. Removal of unsaturated P atoms along the perimeter of the vacancy in the above mentioned order results in the formation of craters aligned along the armchair direction. However, taking a look at the in-plane and the out-of plane bond lengths in bP, the in-plane bond length (2.22 A) is smaller than the out-of-plane bond length (2.24 A). Since smaller bond length is an indicator for a stronger bond, it should be more difficult to desorb P atoms with two in-plane bonds compared to P atoms with one in-plane and one out-of-plane bond. A density functional theory (DFT) calculation considering the desorption of different types of P atoms from the bP surface presented by Fortin-Deschenes et al.~\cite{Fortin-Deschenes2016} also supports this idea by showing that the activation energy of removal of P atoms with one in-plane and one out-of-plane bond (process \# 2 in \cite{Fortin-Deschenes2016}, 3.43~eV) is lower than the activation energy of removal of P atoms with two in-plane bonds (process \#3 in \cite{Fortin-Deschenes2016}, 4.02~eV). Therefore, the order of stability of unsaturated P atoms considered in the model by Liu et al.~\cite{Liu2015} appears unfounded. Performing a similar modeling of a monoatomic P desorption mechanism with the proper order of stability, considering the P atoms with the two in-plane bonds more stably bound than the P atoms with one in-plane and one out-of-plane bond, we obtain craters elongated in zigzag direction (see Fig.~S5 in Supplementary Data).

For the case of the molecular P$_2$ desorption mechanism, Fortin-Deschenes et al.~\cite{Fortin-Deschenes2016} have presented a detailed model, in which the activation energies for various processes of P and  P$_2$ desorption have been estimated by DFT calculations of bP nanoribbons, and then these energies are used in a kinetic Monte Carlo (KMC) simulation of crater formation by sublimation of 10\textsuperscript{6} atoms. These calculations indicate that the craters are aligned along the zigzag direction. This result is substantiated by LEEM measurements reported in Fortin-Deschenes' paper \cite{Fortin-Deschenes2016}, and also agrees well with our results.

Thus, what initially appeared as a very simple criterion to distinguish between the atomic P and the molecular P$_2$ desorption mechanisms, namely if the craters are aligned along the armchair or the zigzag direction, respectively, turns out to be inconclusive based on our careful evaluation of desorption energies, because in either case the craters will be aligned along the zigzag direction. Therefore, measurement of the crater orientation alone is not enough to distinguish between the P and P$_2$ desorption mechanisms. 

Neverthless, Fortin-Deschenes et al.~determined an experimental activation energy for sublimation of $(1.64 \pm 0.10)$~eV from an Arhenius plot \cite{Fortin-Deschenes2016}, in good agreement with their theoretical sublimation activiation energies for P$_2$ desorption. On the other hand, using DFT they report formation energies for divacancies significantly lower than for monovacancies \cite{Fortin-Deschenes2016}, a result later confirmed independently in another paper from a different group \cite{Cai2016}.

\section{Conclusions}

Our study provides information on the annealing conditions (300~$^\circ$C - 350~$^\circ$C) yielding stable and clean bP flakes. It indicates the onset of sample modification at 375~$^\circ$C - 400~$^\circ$C by eye-shaped crater formation due to atomic desorption, and further degradation of the sample at higher temperatures (450~$^\circ$C - 500~$^\circ$C). Furthermore, we examined the craters' preferential long-axis alignment and assigned it to the crystallographic [100] (zigzag) direction. Since, after careful analysis, we concluded that both P and P$_2$ desorption leads to crater elongation along the zigzag direction, the determination of crater alignment alone is not sufficient to pinpoint the underlying desorption mechanism. Evaluation of the energetics of the two mechanisms, as discussed above, gives further insight into the matter. The present is the first surface morphological study of exfoliated few layer bP using STM and provides insight in surface behavior and its degradation with temperature. The latter properties are of much importance in view of the limitations on thermal processing of bP for any practical application of this material.

\section{Methods}

The bP samples were prepared by mechanical exfoliation of bulk bP crystals onto epitaxial monolayer graphene on SiC(0001), which we used as a substrate. The epitaxial monolayer graphene was obtained by annealing 6H-SiC(0001) samples in a high-temperature BM reactor (Aixtron) under Ar atmosphere at about 1400~$^\circ$C and 780~mbar. The bP crystals were prepared by heating commercially-available red phosphorus ($> 99.99$\%) in a muffle oven, together with Sn ($> 99.999$\%), Au ($> 99.99$\%), and a catalytic amount of SnI$_4$, following a published procedure \cite{Nilges2008}. Those solids were loaded into a quartz tube, which was then evacuated by a pumping procedure: the vacuum was backfilled by N$_2$ gas several times, and then the tube was sealed under vacuum. Then it was heated up to 406~$^\circ$C (at a rate of 4.2~$^\circ$C/min), kept 2~h at this temperature, and then heated up to 650~$^\circ$C (2.2~$^\circ$C/min). The sample was kept for three days at this temperature in the oven. Afterwards, a slow cooling rate was chosen (0.1~$^\circ$C/min) to afford the formation of crystals of bP (typical size: 2 mm $\times$ 3 mm).

Exfoliation was carried out with scotch tape inside a glove bag under a constant N$_2$ flux. The bP flakes were then transferred onto monolayer graphene on SiC. In order to limit the exposure of the samples with exfoliated bP to atmosphere,  they were mounted onto the sample holder inside the glove bag, and then transferred to the STM sealed in a plastic bag. The sealing was done also inside the glove bag in N$_2$ environment. Finally, when the load-lock in the STM chamber was ready for sample loading, the plastic seal was broken, the sample was transferred, and load-lock pumping was started immediately, limiting the sample exposure to air to less than a minute.

Once transferred into the preparation chamber of the STM system (base pressure 10\textsuperscript{-10} mbar), the sample was annealed by indirect heating via a BN heater at the backside of the sample. The sample temperature was monitored by a thermocouple and maintained constant with a PID loop. STM measurements were performed at room temperature with an Omicron LT-STM with a base pressure of 10\textsuperscript{-11} mbar. STM tips were prepared by electrochemical etching of tungsten wire with 2M NaOH, and cleaned \textit{in situ} by annealing overnight. Data analysis was performed with WSxM software \cite{Horcas2007}.

\ack

This work was financially supported by EC through the project PHOSFUN {\it Phosphorene functionalization: a new platform for advanced multifunctional materials} (ERC Ad\-vanced Grant No. 670173 to M.~P.), and by CNR--Nano through the SEED project SURPHOS to F.~T.. S.~H.~ thanks Scuola Normale Superiore for support, project SNS16\_B\_HEUN -- 004155. We further acknowledge funding from the Italian Ministry of Foreign Affairs, Direzione Generale per la Promozione del Sistema Paese (agreement on scientific collaboration between Italy and Poland).

\section*{Supplementary Data}

Figures S1--S5 provide additional information on bP flake statistics, surface mor\-phol\-ogy evolution with temperature, statistical analysis of bP surfaces with craters, craters scanned by tip moving along perpendicular directions, and P-monoatomic desorption model resulting in zigzag elongated craters.

\section*{References}

\bibliographystyle{iopart-num}
\bibliography{Abhishek}

\end{document}